\documentclass[prx,floatfix,reprint,amsmath,amssymb,
aps]{revtex4-2}

\usepackage{etoolbox}
\usepackage{graphicx}% Include figure files
\usepackage{dcolumn}% Align table columns on decimal point
\usepackage{bm}% bold math
\usepackage{svg}
\usepackage{booktabs}
\usepackage{multirow}
\usepackage{longtable} 
\usepackage{titlesec}
\usepackage{graphicx}
\titleformat{\section}[hang]{\normalfont\bfseries}{\thesection}{12em}{}
\usepackage{colortbl}
\usepackage{etoolbox}
\makeatletter
\patchcmd{\@makecaption}{\@ifdim{\wd\@tempboxa $>$\hsize}}{\@firstoftwo}{}{}
\makeatother
\usepackage{placeins}  % Add this in preamble
\usepackage{hyperref}
\hypersetup{
    colorlinks=false, % Disable colored links
    hidelinks
}

\begin{document}

%\preprint{APS/123-QED}

\title{Strain engineering of valley-polarized hybrid excitons in a 2D semiconductor}% Force line breaks with \\

\author{Abhijeet M. Kumar$^{1}$}
\author{Douglas J. Bock$^{1}$}
\author{Denis Yagodkin$^{1}$}
% \author{Roberto Rosati$^{2}$}%
\author{Edith Wietek$^{2}$}
\author{Bianca Höfer$^{1}$}
\author{Max Sinner$^{3}$}
\author{Pablo Hernández López$^{4}$}
\author{Sebastian Heeg$^{4}$}
\author{Cornelius Gahl$^{1}$}
\author{Florian Libisch$^{3}$}
\author{Alexey Chernikov$^{2}$}
\author{Ermin Malic$^{5}$}
\author{Roberto Rosati$^{5}$}%
\author{Kirill I. Bolotin$^1$}
\email{kirill.bolotin@fu-berlin.de}
\affiliation{$^1$Department of Physics, Freie Universit{\"a}t Berlin,
Arnimallee 14, 14195 Berlin, Germany}%
\affiliation{$^2$Institute of Applied Physics and Würzburg-Dresden Cluster of Excellence ct.qmat, Technische Universit{\"a}t Dresden, 01062 Dresden, Germany}
\affiliation{$^3$Institute for Theoretical Physics, TU Wien, Wiedner
Hauptstraße 8-10, 1040 Vienna, Austria}
\affiliation{$^4$Department of Physics, Humboldt-Universit{\"a}t Berlin,
Newtonstraße 15, 12489 Berlin, Germany}
\affiliation{$^5$Department of Physics, Philipps-Universit{\"a}t Marburg, 35037 Marburg, Germany}

%TC:ignore

%\affiliation{
%\affiliation{
%\affiliation{%


\begin{abstract}
\textbf{Abstract:} Encoding and manipulating digital information in quantum degrees of freedom is one of the major challenges of today’s science and technology. The valley indices of excitons in transition metal dichalcogenides (TMDs) are well-suited to address this challenge. Here, we demonstrate a new class of strain-tunable, valley-polarized hybrid excitons in monolayer TMDs, comprising a pair of energy-resonant intra- and intervalley excitons. These states combine the advantages of bright intravalley excitons, where the valley index directly couples to light polarization, and dark intervalley excitons, characterized by low depolarization rates. We demonstrate that the hybridized state of dark KK' intervalley and defect-localized excitons exhibits a degree of circular polarization of emitted photons that is three times higher than that of the constituent species. Moreover, a bright KK intravalley and a dark KQ exciton form a coherently coupled hybrid state under energetic resonance, with their valley depolarization dynamics slowed down a hundredfold. Overall, these valley-polarized hybrid excitons with strain-tunable valley character emerge as prime candidates for valleytronic applications in future quantum and information technology.

\end{abstract}

%\keywords{interlayer dark exciton, 2D materials, photocurrent, TMD, lifetime, reflectivity}
%Use showkeys class option if Keyword
\maketitle

%\tableofcontents

%\section{\label{sec:level1}First-level heading:\protect\\ The line
%break was forced \lowercase{via} \textbackslash\textbackslash}
%TC:endignore

\begin{center}
    {\textbf{Introduction}}
\end{center}

The emerging field of valleytronics aims to use the valley index of quasiparticles to store and process quantum information. Transition metal dichalcogenides (TMDs) from the family of layered 2D semiconductors are promising valleytronic materials due to the presence of energy-degenerate extremal points in their band structure (e.g., at K and K' or Q and Q') that host electronic wavefunctions. The K and K' valleys can be selectively addressed by light chirality via valley-contrasting optical selection rules \cite{xiao_coupled_2012,cao_valley-selective_2012}, with the bright energy-degenerate excitons, X$_{\text{KK}}$ and X$_{\text{K'K'}}$ (subscripts denote valley indices of hole and electron wavefunctions, respectively) inheriting the valley index \cite{he_tightly_2014,wang_colloquium_2018,mak_control_2012,lin_high-lying_2022}. 
% Valley polarization of these excitons can be read out by recording the polarization state of the emitted light.
The polarization state of the emitted light provides a direct probe of the valley polarization of these excitons \cite{mak_control_2012}.
Experimental demonstrations of valley magnetization \cite{onga_exciton_2017}, coherent manipulation of valleys \cite{jones_optical_2013,hao_direct_2016}, coupling between valley and spin \cite{xu_spin_2014,mak_control_2012,zeng_valley_2012}, and spatial transport in TMDs \cite{zipfel_exciton_2020,wagner_autoionization_2020,wagner_diffusion_2023,wietek_nonlinear_2024,rosati_negative_2019,tagarelli_electrical_2023} have firmly established the potential of these excitons for valleytronics.

Complex interactions among various intra- and intervalley excitons drive valley depolarization dynamics and transport in TMDs \cite{rosati_dark_2021,zipfel_exciton_2020,rosati_non-equilibrium_2021,dirnberger_quasi-1d_2021,moody_microsecond_2018,plechinger_trion_2016,ersfeld_unveiling_2020}. The ''bright'' intravalley excitons X$_{\text{KK}}$ and X$_{\text{K'K'}}$ suffer rapid valley depolarization due to radiative decay and intervalley exchange interactions on a timescale of picoseconds at cryogenic temperatures \cite{robert_exciton_2016}. These excitons also interact with ''dark'' intervalley excitons, such as X$_{\text{KQ}}$ or X$_{\text{KK'}}$, that form the lowest excited state in tungsten-based TMDs \cite{malic_dark_2018,selig_dark_2018,lindlau_role_2018}. The intervalley excitons weakly couple to light, and are protected from rapid exchange-induced depolarization \cite{selig_ultrafast_2019,selig_suppression_2020}. The spin/valley coupling at the Q valley \cite{beaulieu_berry_2024}, like at the K valley, adds further complexities to the dynamics associated with these states. Finally, momentum-delocalized energy bands slightly below the conduction band, caused by lattice defects, influence valley depolarization by providing the necessary momentum for the radiative recombination of intervalley excitons \cite{linhart_localized_2019,hernandez_lopez_strain_2022}. Despite highly suitable properties of the intervalley excitons for valleytronics, controlling inter-excitonic interactions for the tunable valley polarization and dynamics in TMDs has remained largely unexplored.
% Despite the valley dynamics in TMDs being governed by complex inter-excitonic interactions, controlling these interactions for tunable valley polarization and dynamics has remained largely unexplored.
% Despite these intricate processes governing valley dynamics, controlling inter-excitonic interactions for tunable valley polarization and dynamics has remained largely unexplored.
% These scenarios emphasize the need to tune inter-excitonic interactions and intervalley coupling to enhance control over their valley polarization and dynamics.

\begin{figure*}[t!]
    \includegraphics[width=1\linewidth]{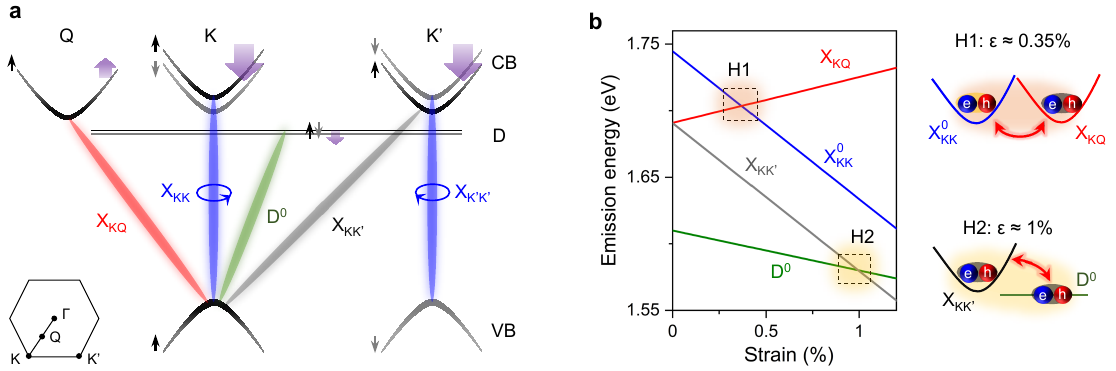}    
    \caption{\textbf{Exciton hybridization in 1L-WSe$_{2}$. a}
    Schematic band structure in 1L-WSe$_{2}$ at zero strain with selected excitonic transitions: X$_{\text{KK}}$ and X$_{\text{K'K'}}$ (blue), X$_{\text{KQ}}$ (red), X$_{\text{KK'}}$ (gray), D$^{0}$ (green). X$_{\text{KK}}$ and X$_{\text{K'K'}}$ transitions couple to light with opposite chirality. The Q/Q’ valleys are also spin/valley locked, similar to the K/K' valleys. The higher energy Q sub-band is not shown here due to a large spin-splitting ($>\,100\,$meV) \cite{zollner_strain-tunable_2019}. Purple arrows denote strain response of valleys with respect to the valence band at the K valley. \textbf{b} Left panel: theoretically calculated strain response of excitons in 1L-WSe$_{2}$, same color-coding as in \textbf{a}. Excitons from different valleys enter energetic resonance near specific strain values, denoted by shaded regions H1 and H2. Note, these calculations do not account for the influence of hybridization on energy shift. Right panel: hybridized state of X$_{\text{KK}}^0$ and X$_{\text{KQ}}$ at $\epsilon \approx 0.35\%$ (H1, top), X$_{\text{KK'}}$ and D$^0$ at $\epsilon \approx 1\%$ (H2, bottom) in excitonic representation.}

    \label{Fig: figure_1}
\end{figure*}

Here, we employ mechanical strain as a powerful tool to modulate intervalley interactions and tune valley polarization dynamics in WSe$_{2}$. Our idea builds on the emerging concept of valley-hybridized excitons, a new class of quasiparticles in TMD monolayers, that we recently demonstrated \cite{hernandez_lopez_strain_2022,kumar_strain_2024}.
The underlying principle involves strain-contrasting energy shift of different TMD valleys \cite{zollner_strain-tunable_2019}, that brings bright and dark excitons with distinct valley character into energetic resonance at specific strain values \cite{khatibi_impact_2018}.
% Crucially, the resulting valley-hybridized excitons present new opportunities beyond conventionally studied bright and dark excitons.
% First, these excitons can inherit the optical properties of the states constituting them, such as large oscillator strength and low intervalley exchange rate.
% Second, exciton hybridization enhances light-matter interactions for the dark excitons, leading to a substantial increase in their optical detectability.
% In contrast, conventional techniques to enhance these interactions are limited to devices without strain control, often requiring methods like magnetic fields \cite{zhang_magnetic_2017} or plasmonics \cite{mueller_photoluminescence_2023}.
% In contrast, strengthening their interactions with light in unstrained devices requires integration of complex techniques such as magnetic fields\cite{zhang_magnetic_2017} or plasmonics\cite{mueller_photoluminescence_2023}.
% , or momentum-resolved photoelectron spectroscopy\cite{noauthor_berry_nodate-3,madeo_directly_2020,wallauer_momentum-resolved_2021}.
% In contrast, dark excitons in unstrained devices are only accessible via advanced techniques such as magneto-, plasmonic-, or momentum-resolved photoelectron spectroscopy techniques.
% Finally, the valley-hybridized excitons are externally tunable by controlling the strain state of the device.
Crucially, valley-hybridized excitons present new means of controlling valley dynamics in TMDs, surpassing conventionally probed bright and dark excitons. First, valley-hybridized excitons can inherit the optical properties of the states constituting them, such as large oscillator strength and reduced intervalley exchange rate.
Second, hybridization enhances light-matter interactions for the dark excitons, leading to a substantial increase in their optical detectability.
This effectively overcomes the usual limitations of conventional techniques --- that are integrated with magnetic fields \cite{zhang_magnetic_2017}, surface plasmon polaritons \cite{zhou_probing_2017}, or waveguides \cite{tang_long_2019} --- in enhancing these interactions in devices without strain control.
% In contrast, conventional techniques to access valley dynamics of dark excitons either require the application of magnetic fields \cite{zhang_magnetic_2017,tang_long_2019}, or rely indirectly on non-radiative interactions with bright excitons \cite{volmer_intervalley_2017}.
% Finally, unlike most conventional approaches to probing valley dynamics, that are restricted to devices without strain control, our straining technique provides external tunability of intervalley interactions and valley-hybridized excitons.
Finally, the valley-hybridized excitons are externally tunable by controlling the strain state of the device.

Leveraging this concept, we focus on the excitons associated with Q valley and momentum-delocalized defect states, which are controllably brought into energetic resonance with K/K' valley excitons in a monolayer of WSe$_{2}$. We find that the emitted photons from these valley-hybridized excitons retain circular polarization, indicating an associated valley index that can be read out via the optical selection rules for K/K' valleys. We examine the states arising from the hybridization between dark intervalley X$_{\text{KK'}}$ and defect-related excitons D$^{0}$, near 0.8\,\% strain. We observe a more than threefold increase in valley polarization of the resulting state, combined with an increase in the radiative recombination rate, compared to the dark excitons in a pristine device. We interpret this as a consequence of the lifting of momentum selection rules upon hybridization. Finally, we investigate the hybridized states of bright X$_{\text{KK}}^{0}$ and dark X$_{\text{KQ}}$ excitons, at $\sim$0.3\,\% strain. We observe a substantial increase in the valley polarization and a remarkable hundredfold slowdown in the depolarization dynamics, which we assign to the suppressed exchange interactions in the resulting coherent state.

\begin{figure*}
    \includegraphics[width=1.0\linewidth]{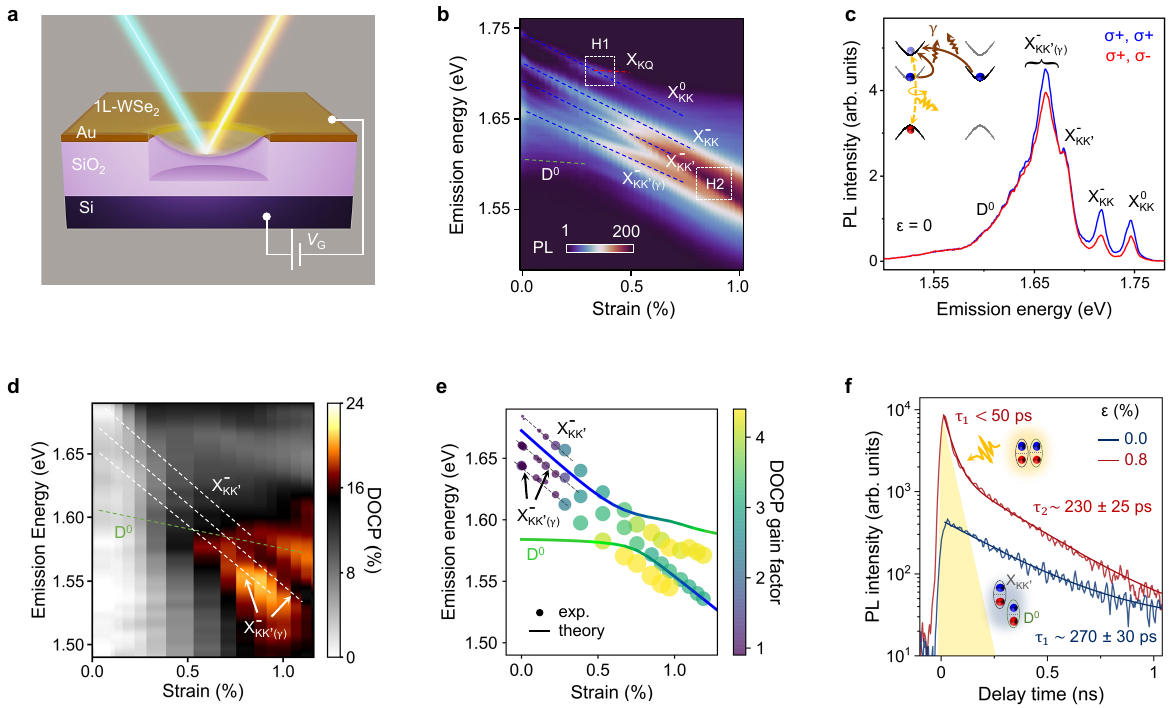}
    \caption{\textbf{Strain tuning of dark K/K’ valley exciton polarization. a} Schematic of the electrostatic straining technique. An applied gate voltage $V_{\text G}$ induces biaxial strain in the center of the membrane. \textbf{b} PL vs. strain false color map in 1L-WSe$_{2}$ (device 1) at $T = 10\,\text K$. The energy shifts of excitons and the hybridization regimes are labeled. \textbf{c} Co- and cross-polarized PL spectra (blue and red, respectively) from 1L-WSe$_{2}$ at $\epsilon = 0$ under $\sigma+$ excitation
    % At $\epsilon = 0$, the device is slightly p-doped, but transitions to n-doping under strain
    (see Note~S4 for discussions on the influence of doping-related effects on $DOCP$). The inset shows radiative emission channels for dark trions X$_{\text{KK'}}^{-}$ mediated by chiral phonons (brown arrows). \textbf{d} $DOCP$ vs. strain false color map in 1L-WSe$_{2}$ (acquired in a different $V_{\text G}$ sweep than in \textbf{b}) for the states below 1.70\,eV. White dashed lines denote the energy shift of X$_{\text{KK'}}^{-}$ and its two phonon replicas; green line extrapolates the same for D$^{0}$. \textbf{e} Scatter plot for the emission energy vs. strain for X$_{\text{KK'}}^{-}$ and its phonon replicas; the area of the circle is proportional to the $DOCP$, the $DOCP$ gain factor ($\eta$) for each state is color-coded. Theoretically calculated strain response of X$_{\text{KK'}}^{-}$ and D$^{0}$ (solid lines), accounting for hybridization effects, reveal an avoided-crossing pattern; color of the line corresponds to the valley character of each exciton (blue: X$_{\text{KK'}}^{-}$, green: D$^{0}$), both lines are downshifted by 15\,meV for clarity. \textbf{f} Time-resolved PL traces in the energetic vicinity of X$_{\text{KK'}}^{-}$ at zero strain (blue) and $0.8\,\%$ strain (red) from the same device. PL intensity at $t = 0$ increases and a new decay component emerges at $0.8\,\%$ strain, highlighted by yellow-shaded region; cartoons depict relative energetic alignment of X$_{\text{KK'}}$ and D$^0$ excitons.}
    \label{Fig: figure_2}
\end{figure*}

\begin{center}
    {\textbf{Results}}
\end{center}

\textbf{Strain-controlled exciton hybridization.} 
% A schematic band structure of an unstrained 1L-WSe$_{2}$ and selected excitonic transitions are shown in Figure~\ref{Fig: figure_1}a. The optically bright neutral excitons X$_{\text{KK}}^{0}$ and X$_{\text{K'K'}}^{0}$ in their respective valleys (yellow in Fig.~\ref{Fig: figure_1}a) are excited by either left- or right-circular polarized light. Intervalley excitons X$_{\text{KK'}}$ and X$_{\text{KQ}}$ (gray in Fig.~\ref{Fig: figure_1}a) reside energetically below X$_{\text{KK}}^{0}$ in a 1L-WSe$_{2}$. An optical transition between the valance band and the momentum-delocalized defect energy bands (D) forms a localized exciton, D$^{0}$. Intervalley excitons X$_{\text{KK'}}$ and X$_{\text{KQ}}$ characterized by weak oscillator strength, require a third particle (e.g., impurity, phonons, etc.) for radiative recombination. The emission intensity of these excitons is influenced by the defect density in the crystal.\\
A schematic band structure of an unstrained 1L-WSe$_{2}$ and selected excitonic transitions are shown in Figure~\ref{Fig: figure_1}a.
The optically bright neutral excitons X$_{\text{KK}}^{0}$ and X$_{\text{K'K'}}^{0}$ (blue in Fig.~\ref{Fig: figure_1}a) reside energetically above the intervalley excitons X$_{\text{KK'}}$ (gray) and X$_{\text{KQ}}$ (red).
% The optically bright neutral excitons X$_{\text{KK}}^{0}$ and X$_{\text{K'K'}}^{0}$ (blue in Fig.~\ref{Fig: figure_1}a) are excited by left- and right-circularly polarized light, respectively. Intervalley excitons X$_{\text{KK'}}$ (gray) and X$_{\text{KQ}}$ (red) reside energetically below X$_{\text{KK}}^{0}$ in a 1L-WSe$_{2}$.
A defect-localized exciton D$^{0}$ (green) is associated with momentum-delocalized energy bands, arising due to point defects, such as single selenium vacancies. The X$_{\text{KK'}}$ and X$_{\text{KQ}}$ excitons are characterized by weak oscillator strengths, and require a third particle (e.g., impurity, phonons, etc.) for radiative recombination \cite{selig_dark_2018,li_momentum-dark_2019,deilmann_finite-momentum_2019,rosati_temporal_2020,brem_phonon-assisted_2020}. As the defects break translational symmetry, radiative recombination of D$^{0}$ is allowed; however, its emission intensity is influenced by the defect density.\\

An applied mechanical strain, $\epsilon$, changes the energetic alignment of valleys, which modifies optical response of all excitonic species \cite{niehues_strain_2018,aslan_strain_2018,harats_dynamics_2020,kovalchuk_neutral_2020}. Under tensile biaxial strain, different valleys exhibit contrasting energy shifts determined by the distinct orbital composition of each valley \cite{feierabend_impact_2017,zollner_strain-tunable_2019}. Specifically, the conduction band (CB) at the K/K' and Q valleys exhibits opposite energy shift with respect to the energies of the valence band (VB) at the K/K' valleys \cite{khatibi_impact_2018,kumar_strain_2024}, while the defect energy bands (D) remain weakly affected (purple arrows in Fig.~\ref{Fig: figure_1}a) \cite{linhart_localized_2019}. Critically, various intra- and intervalley excitons inherit the strain response of the valleys that host their respective electron and hole wavefunctions. Figure~\ref{Fig: figure_1}b shows theoretically calculated strain response of selected excitons in a 1L-WSe$_{2}$. Near 0.35\,\% strain, the dark exciton X$_{\text{KQ}}$ enters energetic resonance with the bright exciton X$_{\text{KK}}^{0}$ (region H1 in Fig.~\ref{Fig: figure_1}b) and acquires oscillator strength \cite{kumar_strain_2024}. Consequently, the resulting hybridized state is optically bright, and shows a distinct optical signature, with its emission energy independent of strain (see Note~S1 for details) \cite{kumar_strain_2024}. Near 1\,\% strain, the intervalley excitons X$_{\text{KK'}}$ hybridize with D$^{0}$ (H2 in Fig.~\ref{Fig: figure_1}b). In this scenario, defect states are predicted to lift momentum selection rules, enabling radiative recombination of the otherwise dark X$_{\text{KK'}}$ excitons \cite{linhart_localized_2019,hernandez_lopez_strain_2022}. Their hybridization is further characterized by an avoided crossing of the excitonic energy levels and a broken spin/valley locking near the K and K' valleys in the CB \cite{linhart_localized_2019}. Overall, the hybridized species acquire traits from the constituting excitons with different valley characters, with the intervalley interactions being strain-dependent. Therefore, we expect exciton hybridization to strongly influence the spin/valley dynamics of excitons in 1L-WSe$_{2}$, with strain playing a key role in controlling these effects.

\textbf{Optical detection of valley-hybridized excitons.}
To induce the mechanical strain required to form valley-hybridized excitons, we employ an electrostatic gating-based straining technique that we recently developed \cite{hernandez_lopez_strain_2022,kumar_strain_2024}. A monolayer of WSe$_{2}$ is suspended over a circular trench of diameter $\sim$5\,\textmu m in an Au/SiO$_{2}$/Si substrate (Fig.~\ref{Fig: figure_2}a). A gate voltage ($V_{\text G}$) applied between WSe$_{2}$ and Si deflects the membrane, inducing tensile biaxial strain in its center, which is symmetric for both polarity of $V_{\text G}$. It is important to note that, although carrier density is inherently linked to the strain state in our approach (see Note~S4 for details), the effects of strain on excitonic energies can be readily distinguished from those originating from carrier density changes \cite{kumar_strain_2024,hernandez_lopez_strain_2022}. We record the excitons' photoluminescence (PL) response vs. strain at $T$ = 10\,K under a CW laser excitation at 1.84\,eV (Fig.~\ref{Fig: figure_2}b; see \textit{Methods} for details). In addition to the well-known bright excitonic states, we identify a series of dark intervalley excitons at $\epsilon$ = 0, including intervalley trions (X$_{\text{KK'}}^{-}$) near 1.685\,eV, their phonon replicas (X$_{\text{KK'($\gamma$)}}^{-}$) around 1.65--1.67\,eV and defect-related excitons near 1.60\,eV by comparing their energetic positions and power dependence with previous reports \cite{hernandez_lopez_strain_2022,rivera_intrinsic_2021,he_valley_2020,kumar_strain_2024}.

Upon applying strain, we observe substantial changes in the PL intensity and emission energy of excitons (Fig.~\ref{Fig: figure_2}b). By tracking the emission energy vs. applied strain and comparing them with the theoretical predictions for the evolution of excitonic energies with strain (Fig.~\ref{Fig: figure_1}b), we identify two hybridization regimes H1 and H2. First, near 0.35\,\% strain, the dark X$_{\text{KQ}}$ becomes energy-resonant with the bright X$_{\text{KK}}^{0}$ exciton and partially brightens due to hybridization (Fig.~S8) \cite{kumar_strain_2024}. Second, near 0.8\,\% strain, the intervalley X$_{\text{KK'}}$ excitons hybridize with D$^{0}$, gaining oscillator strength (Fig.~S4) \cite{hernandez_lopez_strain_2022,linhart_localized_2019}. Consequently, their PL intensity increases by more than an order of magnitude compared to the unstrained state (Fig.~\ref{Fig: figure_2}b). The properties of these hybridized states are detailed in our recent reports \cite{kumar_strain_2024,hernandez_lopez_strain_2022}.

To probe the valley polarization of excitons under strain, we employ polarization-resolved PL measurements under circular or linear excitation. The retention of an exciton's valley memory over its lifetime is quantified by the degree of circular polarization, $DOCP =$ $\frac{I_{\text{co}}-I_{\text{cross}}}{I_{\text{co}}+I_{\text{cross}}}$, where $I_{\text{co}}$ and $I_{\text{cross}}$ are the co- and cross-polarized PL intensities under circularly polarized excitation, respectively. We note that, unlike the case of bright neutral (X$_{\text{KK}}^{0}$) and charged (X$_{\text{KK}}^{+/-}$) excitons, the valley-selective optical selection rules do not directly apply to the intervalley and defect-related transitions. Nevertheless, previous works have demonstrated that the experimentally measured $DOCP$ reliably characterizes their valley polarization state \cite{tang_long_2019,liu_valley-selective_2019}, forming the basis to probe the valley polarization of individual states in our device.
% Therefore, we can probe the valley polarization of individual states in our device by measuring the corresponding $DOCP$.\\

Conversely, the degree of linear polarization, $DOLP = \frac{I_{\text H}-I_{\text V}}{I_{\text H}+I_{\text V}}$, where $I_{\text H}$ and $I_{\text V}$ are the two orthogonal linearly polarized PL components, provides information about the quantum coherence of the entangled states across the two valleys \cite{jones_optical_2013}. The $DOLP$ is non-zero only for X$_{\text{KK}}^{0}$ (Fig.~S8). Valley coherence for dark- and many-body excitons is suppressed due to processes such as intervalley scattering  \cite{jones_optical_2013} and long lifetimes compared to X$_{\text{KK}}^{0}$ \cite{selig_excitonic_2016}, resulting in the vanishing of their $DOLP$.
% Various processes, such as intervalley scattering and long lifetimes of the dark states compared to the X$_{\text{KK}}^{0}$, suppress the valley coherence of many-body excitons, leading to their vanishing $DOLP$ \cite{jones_optical_2013}.

%residing in specific valleys of the electronic band structure.

\textbf{Polarization control of hybridized X$_{\text{KK'}}$--D$^{0}$ excitons.}
Having demonstrated the concept of strain-controlled exciton hybridization, we now investigate the valley-polarized response of K/K' intervalley excitons as they enter energetic resonance with D$^{0}$ (region H2 in Figs.~\ref{Fig: figure_1}b,\,\ref{Fig: figure_2}b). Figure~\ref{Fig: figure_2}c shows co- and cross-circularly polarized PL in 1L-WSe$_{2}$ at zero strain. In this case, X$_{\text{KK'}}^{-}$ excitons show vanishing $DOCP$, consistent with their out-of-plane transition dipoles \cite{wang_-plane_2017,he_valley_2020,li_direct_2019}. A small $DOCP <$ 5\,\% observed from X$_{\text{KK'($\gamma$)}}^{-}$ is likely associated with intervalley scattering processes involving chiral phonons (inset of Fig.~\ref{Fig: figure_2}c) \cite{he_valley_2020}, impurities \cite{rivera_intrinsic_2021}, or resident carriers \cite{robert_spinvalley_2021}. No $DOCP$ is observed from defect-related states near 1.60\,eV.
When strain is applied, the $DOCP$ for these states changes drastically (Fig.~\ref{Fig: figure_2}d). To quantify these changes, we plot the emission energy vs. strain for selected dark states X$_{\text{KK'}}^{-}$ and X$_{\text{KK'($\gamma$)}}^{-}$, obtained using the fitting procedure described in Ref.~\cite{kumar_strain_2024}, in Fig.~\ref{Fig: figure_2}e; the color of the data points reflects the $DOCP$ gain factor $\eta$, defined as $\eta = \frac{DOCP(\epsilon)}{DOCP(\epsilon=0)}$. Note that an experimental uncertainty in $DOCP$ of 5\,\% is assumed to calculate $\eta$ for states with vanishing $DOCP$ at zero strain. 

We make the following observations from the data in Fig.~\ref{Fig: figure_2}e.
First, a maximum in $DOCP$ ($\eta$ between 3 and 5) for dark trions and their phonon replicas is reached at state-specific strain values between 0.7 and 1\,\%, coinciding with the point of hybridization of these states with D$^{0}$. This behaviour also mimics the PL intensity response of the same states, which reaches its maximum in the same strain regime (Fig.~\ref{Fig: figure_2}b). Second, the slope of strain-dependent energy shift for the highest energy dark trions changes for $\epsilon >$ 1\%, aligning with the trend of D$^{0}$ excitons. This suggests that the dark trions acquire, at least partially, the character of D$^{0}$, yet, surprisingly, retain their valley polarization.
This change in the valley character is confirmed by our first principle calculations that show an avoided crossing behaviour between the two states (solid lines in Fig.~\ref{Fig: figure_2}e) by accounting for strain-dependent coupling between them (see Note~S2, Fig.~S3 for details).
% Our first principle calculations for strain-dependent energy shifts (solid lines in Fig.~\ref{Fig: figure_2}e), accounting for the coupling between the two states, confirm an avoided crossing behaviour and a shift in the energy trend near 1\% strain.
% Consequently, D$^0$ excitons retain valley polarization up to the maximum applied strain level of 1.8\,\% (Fig.~S4).

Finally, we find that the temporal dynamics of the dark trions, recorded via time-resolved PL (see \textit{Methods} for details), change significantly near the hybridization point (Fig.~\ref{Fig: figure_2}f). At $\epsilon$ = 0.8\,\%, the PL intensity at $t$ = 0 increases tenfold, accompanied by the emergence of new decay component ($\tau_1 < $ 50\,ps), indicating a similar increase in the recombination rate compared to the zero strain case. This directly reflects the increase of the light-matter coupling strengths due to exciton hybridization and demonstrates the associated increase in the radiative recombination rate.
% absence of additional strain-induced non-radiative channels.

% The data in Figs.~\ref{Fig: figure_2}d-f collectively highlight the influence of inter-excitonic hybridization on the valley polarization of dark K/K' valley excitons. The amplification in $DOCP$ is concomitant with a maximum in PL intensity, reduced radiative lifetime and an avoided crossing between the defect-bound and the dark K/K' valley excitons. The hybridization lifts momentum-selection rules, driving radiative recombination of the dark excitons and enabling the detection of their valley polarization. A positive sign of the $DOCP$ further confirms that the photons are emitted from the valley of excitation, suggesting a preferred intervalley scattering of electrons and the valley polarization state of the photoexcited holes. Finally, we note that while the straining technique also changes the carrier density in these devices, such effects do not explain the observed amplification in $DOCP$ (see Note~S4, Figs.~S5,S6 for details).

The data in Figs.~\ref{Fig: figure_2}d-f collectively highlight the influence of inter-excitonic hybridization on the valley polarization of dark K/K' valley excitons. The amplification in $DOCP$ is concomitant with a maximum in PL intensity, reduced radiative lifetime, and an avoided crossing between the defect-bound and the dark K/K' valley excitons. This is consistent with resonant coupling with defects relaxing the momentum-selection rules, driving radiative recombination of dark excitons and enabling direct optical detection of their valley polarization. The faster recombination prevents the mixing of valley states in the hybridized species, leading to an enhanced $DOCP$. The valley polarization is further strengthened by the hole's valley state in the VB, which predominantly resides in the valley of excitation. Notably, the D$^0$ exciton retains valley polarization up to the maximum applied strain level of 1.8\,\% (Fig.~S4), indicating that this state has inherited the valley character of ’free’ K/K’ valley excitons after hybridization. Finally, we note that while the straining technique also changes the carrier density in these devices, such effects do not explain the observed amplification in $DOCP$ (see Note~S4 and Figs.~S5,\,S6 for details).

\begin{figure*}
    \includegraphics[width=0.8\linewidth]{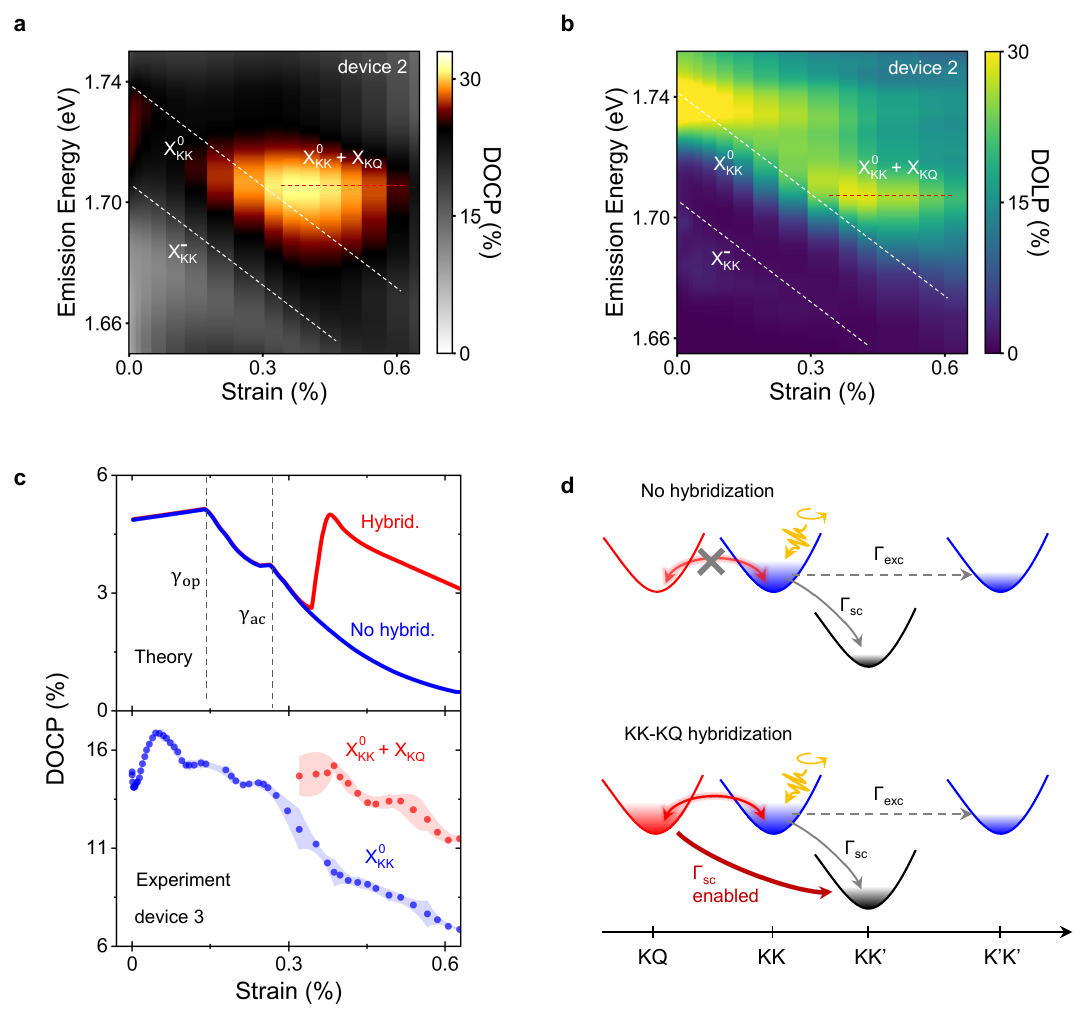} 
    \caption{\textbf{Valley polarization control by tuning KK--KQ coupling. a,b} False color map of $DOCP$ (\textbf{a}) and $DOLP$ (\textbf{b}) vs. strain from device 2. A local maximum in $DOCP$ and $DOLP$ is observed near 0.4\,\% strain, suggesting an enhanced intervalley coupling and the formation of a coherent hybridized state. Note, the nonzero polarization on the higher energy tail arises from excitons located away from the membrane center due to strain inhomogeneity. 
    % of X$_{\text{KK}}^{0}$ due to low PL intensity and contributions from excitons away from the center of the membrane due to strain inhomogeneity. 
    \textbf{c} Top panel: theoretically calculated $DOCP$ vs. strain for X$_{\text{KK}}^{0}$ without (blue) and with (red) hybridization with X$_{\text{KQ}}$. The two dashed lines denote strain values corresponding to the closing of optical ($\gamma_{\text{op}}$)- and acoustic phonon ($\gamma_{\text{ac}}$)-mediated scattering channel between X$_{\text{KK}}^{0}$ and X$_{\text{KQ}}$. Bottom panel: experimentally obtained $DOCP$ vs. strain (from device 3). An increase in $DOCP$ near 0.4\,\% strain is observed for the hybridized state (red), showing qualitative agreement with the theory.
    The $DOCP$ trend below 0.1\% strain is not yet understood. 
    \textbf{d} Schematic representation of intervalley exciton scattering near $\epsilon = 0.35\,\%$. X$_{\text{KK}}^{0}$, X$_{\text{K'K'}}^{0}$ and X$_{\text{KQ}}$ are energy-resonant, lying above X$_{\text{KK'}}$. Yellow arrows denote the valley polarized excitation of KK excitons, whereas gray arrows denote intervalley exchange (dashed) and phonon-assisted scattering (solid). In absence of hybridization (top), phonon-assisted scattering between X$_{\text{KK}}^{0}$ and X$_{\text{KQ}}$ is blocked. The hybridization (bottom) forms a resonantly coupled KK--KQ state and enables a new scattering channel between X$_{\text{KQ}}$ and X$_{\text{KK'}}$ (dark red arrow).}
    \label{Fig: figure_3}
\end{figure*}

\textbf{Polarization control of hybridized X$_{\text{KK}}^{0}$--X$_{\text{KQ}}$ excitons.}
Next, we focus on the hybridized state of X$_{\text{KQ}}$ and X$_{\text{KK}}^{0}$ excitons (region H1 in Figs.~\ref{Fig: figure_1}b,\,\ref{Fig: figure_2}b), which is characterized by a new peak in the PL spectrum near 1.70\,eV and near 0.35\,\% strain (Fig.~S8) \cite{kumar_strain_2024}. Both the strain range and the spectral window of this hybridization are different than the previously discussed regime H2, corresponding to the X$_{\text{KK'}}$--D$^{0}$ hybridization.

Figures 3a and 3b show false color maps of $DOCP$ and $DOLP$ vs. strain, respectively, in the energetic vicinity of X$_{\text{KK}}^{0}$. We detect an increase in both $DOCP$ and $DOLP$ in the regime corresponding to the formation of the hybridized state. Since the retention of linear polarization is a measure of exciton valley coherence, the observation of an increased $DOLP$ suggests that the energy-resonant X$_{\text{KK}}^{0}$ and X$_{\text{KQ}}$ have formed a coherently-coupled hybrid state. 
% This further supports the conditions for the polarization increase, taking place only in a restricted strain range \cite{kumar_strain_2024}.
% and at a fixed energy in analogy to the recently observed X$_{\text{KQ}}$ peak \cite{kumar_strain_2024}.
% This can be traced back to the fact that  the hybridization between KQ and KK excitons occurs  only when these two states  are quasi-resonant, hence only in a restricted strain range  and at a given energy [28].
An increased $DOCP$ also suggests the coupling between the valley indices of the excitons from K and Q valleys, which has remained unexplored until now.
% The influence of K-Q hybridization on exciton valley polarization has remained unexplored until now.
% However, so far it has remained in the dark, if the hybridization can amplify the polarization.  

To explain our observations, we develop a many-particle theory to microscopically determine the strain-dependent exciton landscape, both with and without exciton hybridization. Starting from exciton dynamics in the presence of exchange interactions \cite{selig_suppression_2020} and doping \cite{yu_dirac_2014}, we include the continuous-wave excitation to obtain the relative occupation of the bright excitons with opposite circular polarization (see \textit{Methods} and Note~S1). This allows us to write the $DOCP$ as

\begin{equation}
    DOCP = \frac{\Gamma_{\text{dec}}}{2\Gamma_{\text{exc}} + \Gamma_{\text{dec}}},
    \label{eq: Rate_eq}
\end{equation}

where $\Gamma_{\text{dec}} = \Gamma_{\text{rad}} + \Gamma_{\text{sc}}$ represents the combined rates associated with the radiative decay of excitons ($\Gamma_{\text{rad}}$) and phonon-assisted intervalley scattering ($\Gamma_{\text{sc}}$), while excluding the intervalley exchange rate $\Gamma_{\text{exc}}$.
% the total decay rate excluding the intervalley exchange interactions, combining the radiative decay rate of excitons ($\Gamma_{\text{rad}}$) and phonon-assisted intervalley scattering rates ($\Gamma_{\text{sc}}$).
% , with contributions from both radiative processes and phonon-assisted scattering with rate $\Gamma_{\text{sc}}$
% The intervalley exchange rate $\Gamma_{\text{exc}}$
In the absence of hybridization, the latter decay rate is microscopically evaluated as $\Gamma_{\text{exc}} = \frac{2}{\Gamma_{\text{sc}}}J^{2}$, where $J^{2} = \Sigma_{\bm{q}}|J_{\bm{q}}|^{2}N^{\circ}_{\bm{q}}$ is the squared modulus of the exchange Hamiltonian $|J_{\bm{q}}|^{2}$, summed over the exciton momentum $\bm{q}$ and weighted by the normalized excitonic distribution, here assumed to be thermalized (Note~S1). In agreement with the generalized Maialle-Silva-Sham model \cite{wu_enhancement_2021,maialle_exciton_1993}, the derived  $\Gamma_{\text{exc}}$  is quadratic in the exchange Hamiltonian --- which depends weakly on strain --- and scales with the inverse of the scattering rates $\Gamma_{\text{sc}}$.
% , which furthermore provides the main contribution to the decay rates $\Gamma_{\text{dec}}$  together with the radiative rates.
We further evaluate $\Gamma_{\text{sc}}$ microscopically, revealing its crucial dependence on strain and hybridization.
% In general, the $DOCP$ of X_${\text{KK}}^{0}$, as described by Eq.~\eqref{eq: Rate_eq}, is defined by the competition between $\Gamma{\text{exc}}$ and $\Gamma_{\text{sc}}$, in complete agreement with the independently employed rate equation model.
Our final calculations for the $DOCP$ of X$_{\text{KK}}^{0}$, via Eq.~\eqref{eq: Rate_eq}, show strong strain-dependent behaviour (Fig.~\ref{Fig: figure_3}c, top panel). Note, that the evaluation of $DOCP$ via Eq.~\eqref{eq: Rate_eq} is consistent with an independently employed rate equation model \cite{mak_control_2012,an_strain_2023}.

% Correspondingly, Eq.~\eqref{eq: Rate_eq}, defining the $DOCP$ of  X$_{\text{KK}}^{0}$  as a competition between $\Gamma_{\text{dec}}$ and $\Gamma_{\text{exc}}$ shows strong strain-dependent behaviour.
In the absence of hybridization, our microscopic calculations show a decrease of $DOCP$ with increased strain, with two pronounced drops near 0.15\,\% and 0.27\,\% strain (dashed lines in Fig.~\ref{Fig: figure_3}c, top).
These features reflect the strain-induced closing of optical and acoustic phonon-assisted scattering channels between X$_{\text{KK}}^{0}$ and X$_{\text{KQ}}$ (Note~S1).  Consequently, the decay rate $\Gamma_{\text{dec}}$ and, hence, the $DOCP$ lowers.  Notably, the $DOCP$ increases sharply near $\epsilon$ = 0.35\,\% when the hybridization between X$_{\text{KK}}^{0}$  and X$_{\text{KQ}}$ is accounted for in the model (red in Fig.~\ref{Fig: figure_3}c, top). 
% Thanks to this hybridization, the bright states acquire a KQ component along with the associated phonon-assisted scattering channel from X$_{\text{KQ}}$ to X$_{\text{KK'}}$, which is highly efficient, as demonstrated by strain-dependent diffusion studies[54].
Thanks to this hybridization, the bright states acquire a KQ component and, subsequently, gain the scattering channel from KQ to KK'. This scattering mechanism is particularly effective in this strain regime, as demonstrated recently by strain-dependent diffusion studies \cite{rosati_strain-dependent_2020}.
In contrast, the scattering between X$_{\text{KK}}^{0}$ and X$_{\text{KK'}}$ changes only weakly with strain due to their similar strain-dependent energy shifts (Fig.~S2). We also assume negligible changes in the radiative decay rate $\Gamma_{\text{rad}}$ of X$_{\text{KK}}^{0}$ as its binding energy remains nearly unaffected in the studied strain range \cite{kumar_strain_2024}. Consequently, this new scattering channel between the hybridized states and X$_{\text{KK'}}$ leads to the increase of $\Gamma_{\text{dec}}$ (depicted by the cartoon in Fig.~\ref{Fig: figure_3}d), and with it, to a local maximum in $DOCP$ at $\sim$0.4\,\% strain.

Our experimental observations in Fig.~\ref{Fig: figure_3}c (bottom panel) are in qualitative agreement with the theoretical predictions. Furthermore, our theory also explains a gradual decrease of $DOCP$ vs. strain for the hybridized state in experiments, which we ascribe to the residual inhomogeneity of the strain profile in our device (see Note~S1, Fig.~S1 for details).\\
% Differences between theory and experiment including the non-monotonic behaviour in $<$ 0.1\,\%   are discussed in the Supplemental Material.\\

\begin{figure}[t!]
    \includegraphics[width=1\linewidth]{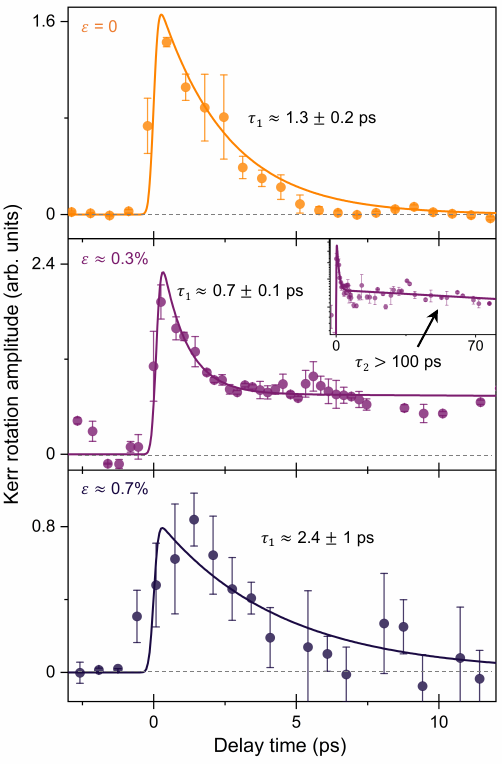}    
    \caption{\textbf{Ultrafast spin/valley dynamics vs. strain.} Top-bottom: TRKR dynamics in 1L-WSe$_{2}$ (device~2) at $\epsilon =$ 0, 0.3 and 0.7\,\%, respectively, at $T = 100\,\text K$. The pump and probe energies (E$_{\text{pump}}$ = E$_{\text{probe}}$) were set near the neutral exciton resonance for each strain value (1.730, 1.695, and 1.650\,eV, respectively), determined via \textit{in-situ} PL measurements.
    % The corresponding pump/probe energies (E$_{\text{pump}}$ = E$_{\text{probe}}$) are labeled.
    Solid lines are fits, dashed gray lines correspond to zero TRKR amplitude. The inset in the middle panel show the TRKR trace at $\epsilon =$ 0.3\,\% on a longer timescale in a semi-log plot. A second decay component with a time constant exceeding 100\,ps is observed.
    % The TRKR signal decays mono-exponentially within 2\,ps of photoexcitation at $\epsilon =$ 0. In contrast, the TRKR dynamics show significant changes near 0.3\% strain as it turns bi-exponential and a second decay component emerges with a time constant more than two orders of magnitude larger compared to $\epsilon = 0$. At $\epsilon = 0.3\%$, a rapid spin/valley depolarization is observed.
    }
    \label{Fig: figure_4}
\end{figure}

\textbf{Ultrafast dynamics of hybridized X$_{\text{KK}}^{0}$--X$_{\text{KQ}}$ excitons.}
Finally, our straining technique provides direct access to the investigation of spin/valley dynamics stemming from the resonant coupling between K and Q valley excitons. To this end, we employed single-color time-resolved Kerr rotation (TRKR) spectroscopy. Here, a circularly polarized pump pulse initializes a valley-polarized optical transition near the X$_{\text{KK}}^{0}$ resonance, while the spin/valley imbalance is probed by a time-delayed linearly polarized probe pulse (see \textit{Methods} and Fig.~S9 for details). We chose the temperature $T = 100\,$K to 
minimize the contributions from dark excitons \cite{plechinger_trion_2016,volmer_intervalley_2017}, resident carriers \cite{ersfeld_unveiling_2020,yang_long-lived_2015,dey_gate-controlled_2017}, and trap states \cite{li_valley_2021,ersfeld_unveiling_2020}, that are dominant at cryogenic temperatures and complicate the interpretation of the data.

Figure~\ref{Fig: figure_4} shows TRKR traces at selected strain values; solid lines are the fits to the data. We observe a rapid decay of the TRKR signal at zero strain, with a time constant smaller than 2\,ps (top panel). The TRKR dynamics change significantly near $\epsilon = 0.3\,\%$, the value corresponding to X$_{\text{KK}}^{0}$--X$_{\text{KQ}}$ resonance (middle panel). Here, the decay dynamics turn bi-exponential, with the 
second decay component ($\tau_2$) longer-lived compared to the unstrained case by more than two orders of magnitude. As the applied strain is increased to 0.7\,\%, the TRKR signal decay accelerates again, becoming qualitatively similar to the unstrained case, with a single decay constant of 2.4$\,\pm\,1\,$ps (bottom panel).

A short-lived TRKR dynamics at zero strain can be explained by rapid exchange-mediated intervalley scattering --- expected to be dominant under resonant pump excitation --- and phonon-mediated valley depolarization. We ascribe the long-lived TRKR dynamics near $\epsilon = 0.3\,\%$ to KK--KQ hybridization. Here, the KK state acquires a partially intervalley character, for which the exchange scattering mechanism is suppressed \cite{selig_suppression_2020}. 
Correspondingly, the second decay component reflects a low rate of valley depolarization for the hybridized population. This assignment is further supported by a similar trend consistently recorded across devices with varying doping levels, when probed near the X$_{\text{KK}}^{0}$ resonance and near 0.3\,\% strain (Figs.~S11,\,S12). The TRKR dynamics probed near the X$_{\text{KK}}^{+/-}$ resonance do not exhibit any long-lived component, with their decay constants ranging from 2 to 6\,ps across the applied gate voltage range (Fig.~S10). These findings rule out the the dynamics near $\epsilon = 0.3\,\%$ being governed by doping-related resident carrier depolarization \cite{dey_gate-controlled_2017,ersfeld_unveiling_2020,yang_long-lived_2015}. Furthermore, the spin/valley dynamics in our devices are free from substrate-related disorder, which are typically linked to doping-dependent transitions between mono- and bi-exponential decay profiles \cite{li_valley_2021}. Finally, the disappearance of the long-lived component in the TRKR dynamics at 0.7\,\% strain (bottom panel in Fig.~\ref{Fig: figure_4}) is consistent with X$_{\text{KK}}^{0}$ and X$_{\text{KQ}}$ coming out of energetic resonance.
Overall, our observations in Fig.~\ref{Fig: figure_4} demonstrate the immense potential of strain in tuning ultrafast spin/valley dynamics in TMDs.
\FloatBarrier

\begin{center}
    {\textbf{Discussion}}
\end{center}
To summarize, we demonstrated the emergence of valley-polarized hybrid excitons in WSe$_{2}$ monolayers under controlled strain. By employing polarization-resolved spectroscopy,
% and TRKR microscopy, we explored changes in the valley character and spin/valley polarized responses of these excitons.
we recorded a more than threefold increase in the valley polarization of dark K/K' valley excitons upon hybridization with defect-related excitons. The hybridization with the KQ intervalley excitons increases the valley polarization of X$_{\text{KK}}^{0}$ excitons and reduces their spin/valley depolarization rates by more than two orders of magnitude.
% These results are first spectroscopic demonstration of the influence of the coupling between K and Q valleys on exciton valley polarization and dynamics.

Our findings include a first spectroscopic demonstration of strain-controlled intervalley coupling on valley polarization dynamics, and open multiple directions for future research. First, it may be possible to achieve near 100\,\% valley polarization of hybridized excitons in future experiments with tunable, near-resonant excitation. We highlight that the reported valley polarization in our PL experiments is underestimated by up to 50\,\% due to strain-induced energy detuning under fixed excitation energy (Fig.~S7). Long lifetimes of species associated with the Q valley may be utilized for spin/valley-polarized transport, achievable in devices with engineered strain gradients.
% , featuring a ''source'' region for optical excitation and ''drain'' for readout.
% This is because, in these experiments, the excitation energy is fixed while the energy positions of the excitonic peaks are strain-dependent. As a result, there is a strain-dependent decrease in the $DOCP$ due to a continuous reduction in the polarization of the initially photoexcited KK excitons. and could be much higher under tunable, near-resonant excitation.
% Second, our results highlight the potential of long-lived, spin/valley-coupled X$_{\text{KQ}}$ excitons for valleytronic applications. We propose a device with engineered strain gradients to drive valley-polarized transport and subsequent optical readout.
Second, our results opens new paths for theoretical exploration, particularly in understanding the microscopic interactions between excitons and phonons in the presence of defects, as well as the spin/valley depolarization mechanisms for the hybridized KK--KQ excitons.
% Second, g-factor measurements will help resolve the fine structure of valley-hybridized excitons. Extended theory models will improve the microscopic understanding of
% coherently-coupled X$_{\text{KK}}^{0}$ and X$_{\text{KQ}}$ excitons. E
% exciton-phonon interactions in the presence of defects and the spin/valley depolarization mechanisms for the hybridized KK-KQ excitons.
Finally, an exciting prospect is to describe the valley index in the language of pseudospin, 
% where the coherent superposition of excitonic states in two energy-degenerate valleys is mapped onto a $\frac{1}{2}$ spinor. In this framework,
where the effect of uniaxial strain is equivalent to that of a magnetic field acting in the pseudospin space \cite{yu_dirac_2014,glazov_exciton_2022,iakovlev_fermi_2023}. We recently demonstrated this pseudomagnetic field exceeding 40 Tesla in uniaxially-strained TMDs, enabling the study of analogs of the Zeeman and Larmor effects, as well as tunable exciton coherence dynamics \cite{yagodkin_excitons_2024}. Investigating the coherent nature of KK--KQ excitons under pseudomagnetic fields \cite{thompson_anisotropic_2022} would be an interesting next step towards valley manipulation.\\

{\textbf{Methods}}

\textit{Sample fabrication}
The WSe$_2$ flakes were mechanically exfoliated and transferred onto a circular trench (diameter is $\sim$5\,\textmu m) in a Au/Cr/SiO$_2$/Si stack using a dry transfer approach. The cavity was developed via wet etching process using Hydrofluoric (HF) acid. A gate voltage (typically
in the range of up to $\pm210\,$V) was applied between the TMD flake (electrically
grounded) and the Si back gate of the chip to induce strain. The strain in the center was
characterized following the laser interferometry approach used in our recent work \cite{kumar_strain_2024}.

\textit{Polarization-resolved PL measurements}
The devices were measured inside a cryostat (CryoVac Konti Micro) at a base temperature of 10\,K. Devices~1~and~2 (corresponding to data in Figs.~\ref{Fig: figure_2} and ~\ref{Fig: figure_3}a,b) were probed under a CW laser excitation at $\lambda = 670\,$nm (6\,\textmu W), tightly focused in the center of the membrane with spot diameter $\sim$1\,\textmu m. For device 3 (data in Fig.~\ref{Fig: figure_3}c), we used a CW laser with $\lambda = 685\,$nm and 2\,\textmu W power.
A combination of a polarizer (GL\,10, Thorlabs) and a quarter-wave plate (RAC\,4.4.10, B. Halle) was used to control the circular polarization of excitation. For linearly-polarized excitation, the quarter-wave plate was replaced with a half-wave plate (RAC\,4.2.10, B. Halle). Both wave plates were positioned before the objective, ensuring that the emitted light passed through them before getting detected. To filter the polarization state of the emitted light, we used a combination of a half-wave plate and an analyzer before the spectrometer. The PL signal was detected using the Spectrometer Kymera 193i Spectrograph.

\textit{Time-resolved PL measurements}
The sample was excited by a pulsed Ti:sapphire laser (Coherent Chameleon Ultra II, $140\,$fs pulse duration and 80\,MHz repetition rate) at $\lambda = 700\,$nm and 0.55\,\textmu Jcm$^{-2}$ fluence, with a laser spot of $\sim$1\,\textmu m diameter, focused onto the sample using a 60x objective. The injected electron-hole pair density was on the order of 10$^{10}$\,cm$^{-2}$, where Auger-like exciton-exciton annihilation effects are negligible. A streak camera (C10910, Hamamatsu) was used to time- and spectrally resolve the PL signal.

\textit{Time-resolved Kerr rotation microscopy}
We used a wavelength-tunable pulsed Ti:sapphire laser (Coherent Chameleon Ultra II, $140\,$fs pulse duration and 80\,MHz repetition rate). The laser pulse was split into the pump and probe components. Circular polarization of the pump pulse was controlled using a combination of a linear polarizer and a quarter wave plate. Both the pulses were made collinear while being spatially separated, and were focused onto the sample at near normal incidence using a reflective objective (LMM40X-P01, Thorlabs). The pump and probe spot sizes were $\sim$3 and $\sim$1\,\textmu m, respectively. The reflected pump beam was blocked using a spatial filter. The reflected linearly-polarized probe was separated into two components of orthogonal polarization after passing through an optical bridge consisting of a half-wave plate and a Wollaston prism (Fig.~S9). Both polarization components were simultaneously recorded by using a home-built balanced photodetector, and the differential signal was read out using a lock-in amplifier phase-locked to a chopper in the pump path. For all TRKR measurements, two traces with opposite pump helicity were recorded sequentially, and the average of these traces was calculated. The strain level and corresponding exciton resonances during the TRKR measurements were characterized by \textit{in-situ} PL (Fig.~S9).

\textit{Microscopic many-particle modeling}
We calculate  exciton energies by solving a strain-dependent Wannier equation \cite{rosati_strain-dependent_2020,kumar_strain_2024}, 
% [2D Materials 8, 015030 (2021); Ref. [29]]
including the Keldysh-Rytova Coulomb potential \cite{rytova_screening_1967,keldysh_coulomb_1979}, 
% [L. Keldysh, Jetp Letters (1979).]
and single-particle valley-dependent masses and strain-induced energy shifts \cite{kormanyos_kp_2015,khatibi_impact_2018,zollner_strain-tunable_2019}. 
% [2D Materials 2, 022001 (2015); 2D Materials 6, 015015 (2018); Physical Review B 100, 195126 (2019).]
The resulting excitonic eigenstates allow to evaluate the corresponding valley-dependent exciton-phonon scattering rates \cite{rosati_strain-dependent_2020}. 
% [2D Materials 8, 015030 (2021)]
Assuming that hybridization occurs for quasi-resonant KK and KQ states, we are able to predict  experimentally measured  strain-dependent PL signatures. In order to predict the $DOCP$, we evaluate the relative occupation of  bright exciton states with opposite circular polarization. For this purpose, we start from exciton dynamics in the presence of exchange interaction \cite{selig_suppression_2020} and doping \cite{yu_dirac_2014} and include the continuous-wave excitation. More details can be found in the Supplementary Information (Note~S1).

\textbf{Acknowledgments}

The research group from Freie Universit{\"a}t Berlin acknowledges the Deutsche Forschungsgemeinschaft (DFG) for financial support through the Collaborative Research Center TRR 227 Ultrafast Spin Dynamics (project B08) and the Priority Programme SPP 2244 2DMP (project BO 5142/5) as well as the Federal Ministry of Education and Research (BMBF, Projekt 05K22KE3).
The Marburg group acknowledges financial support by the DFG via SFB 1083 (project B9) and the regular project 512604469.
The Dresden group acknowledges financial support by the DFG via SPP2244 (CH 1672/3, Project-ID: 443405595), joint project with the Marburg team (Project-ID: 287022282), SFB 1277 (Project B05, Project-ID No. 314695032), ERC CoG CoulENGINE (GA number 101001764) and Würzburg-Dresden Cluster of Excellence on Complexity and Topology in Quantum Matter ct.qmat (EXC 2147, Project-ID 390858490).
Research by the Vienna group has been supported by The Austrian Science Fund (FWF) through doctoral college TU-DX (DOC 142-N) and the MECS cluster of excellence 10.55776/COE5 MECS. For the purpose of open access, the author has applied a CC BY public copyright licence to any Author Accepted Manuscript version arising from this submission.
The Humboldt-Universit{\"a}t Berlin group acknowledges funding from the DFG under the Emmy Noether Initiative (project-ID 433878606).
We thank N. Stetzuhn and A. Dewambrechies for useful comments.

\textbf{Author Contributions}

A.M.K. and K.I.B. conceived the project. A.M.K., C.G., and D.Y. designed the experimental setup. A.M.K., D.J.B., B.H., and P.H.L. prepared the samples. S.H. and P.H.L. developed the electrostatic straining technique. A.M.K., D.J.B., and D.Y. performed polarization-resolved PL and TRKR measurements. E.W. and A.C. performed time-resolved PL measurements. R.R. and E.M. developed theory for KQ excitons. M.S. and F.L. developed the theory for
defect excitons. A.M.K., D.B., E.W., and D.Y. analyzed the data. A.M.K., K.I.B., and R.R. wrote the manuscript with input from all co-authors.

\textbf{Data Availability Statement}

The data that support the findings of this study are available from the corresponding author upon reasonable request.

The authors declare no competing interest.

\bibliography{main}
% \input{output.bbl}

%TC:endignore
\end{document}